\documentclass[12pt]{article}
\usepackage{amssymb}
\begin{document}
\begin{center}
In-medium QCD 
and Cherenkov gluons vs Mach waves at LHC \\

I.M. Dremin\\ 

{\it Lebedev Physical Institute, Moscow, Russia}

\end{center}

\begin{abstract}
The equations of in-medium gluodynamics are proposed. Their 
classical lowest order solution is explicitly shown for a color charge
moving with constant speed. For nuclear permittivity larger than 1 it
describes the shock wave induced by emission of Cherenkov gluons. Specific 
effects at LHC energies are described and compared with Mach wave predictions.
\end{abstract}

The properties and evolution of the medium formed in ultrarelativistic heavy-ion 
collisions are widely debated. At the simplest level it is assumed to consist 
of a set of current quarks and gluons. The collective excitation modes of the 
medium may, however, play a crucial role.
One of the ways to gain more knowledge about the excitation modes is to 
consider the propagation of relativistic partons through this matter.
Phenomenologically their impact would be described by the nuclear permittivity
of the matter corresponding to its response to passing partons. Namely this
approach is most successful for electrodynamical processes in matter.
Therefore, it is reasonable to modify the QCD equations by taking into account
collective properties of the quark-gluon medium \cite{depj}. Strangely enough,
this was not done earlier. For the sake of simplicity we consider here the 
gluodynamics only.

The classical lowest order solution of these equations coincides with
Abelian electrodynamical results up to a trivial color factor. One of the most
spectacular of them is Cherenkov radiation and its properties. Now, Cherenkov
gluons take the place of Cherenkov photons \cite{d1, ko}. Their emission in 
high-energy hadronic collisions is described by the same formulae but 
with the nuclear permittivity in place of the usual one. Actually, one 
considers them as quasiparticles, i.e. quanta of the medium excitations 
leading to shock waves with properties determined by the permittivity. 

Another problem of this approach is related to the notion of the
rest system of the medium. It results in some specific features of this
effect at LHC energies.

To begin, let us recall the classical in-vacuum Yang-Mills equations
\begin{equation}
\label{f.1}
D_{\mu}F^{\mu \nu }=J^{\nu }, \;\;\;\;\;
F^{\mu \nu }=\partial ^{\mu }A^{\nu }-\partial ^{\nu }A^{\mu }-
ig[A^{\mu },A^{\nu }],
\end{equation}
where $A^{\mu}=iA_a^{\mu}T_a; \; A_a (A_a^0\equiv \Phi_a, {\bf A}_a)$ are the 
gauge field (scalar and vector) potentials, the color matrices $T_a$ satisfy
the relation $[T_a, T_b]=if_{abc}T_c$, $\; D_{\mu }=\partial _{\mu }-ig[A_{\mu }, \cdot], \;\; 
J^{\nu }(\rho, {\bf j})$ a classical source current, the 
metric $g^{\mu \nu }$=diag(+,--,--,--).

In the covariant gauge $\partial _{\mu }A^{\mu }=0$ they are written  
\begin{equation}
\label{f.2}
\square A^{\mu }=J^{\mu }+ig[A_{\nu }, \partial ^{\nu }A^{\mu }+F^{\mu \nu }],
\end{equation}
where $\square $ is the d'Alembertian operator. 

The chromoelectric and chromomagnetic fields are
$E^{\mu}=F^{\mu 0 }, \;
B^{\mu}=-\frac {1}{2}\epsilon ^{\mu ij}F^{ij}$
or, as functions of the gauge potentials in vector notation,
\begin{equation}
\label{4}
{\bf E}_a=-{\rm grad }\Phi  _a-\frac {\partial {\bf A}_a}{\partial t}+
gf_{abc}{\bf A}_b \Phi _c, \;\;\;\;
{\bf B}_a={\rm curl }{\bf A}_a-\frac {1}{2}gf_{abc}[{\bf A}_b{\bf A}_c].
\end{equation}

Herefrom, one easily rewrites the in-vacuum equations of motion (\ref{f.1}) in 
vector form. We do not show them explicitly here (see \cite{depj}) and write 
down the equations of the in-medium gluodynamics using the same method 
as in electrodynamics. We introduce the nuclear
permittivity and denote it also by $\epsilon $, since this will not lead
to any confusion. After that, one should replace ${\bf E}_a$ 
by $\epsilon {\bf E}_a$ and get
\begin{equation}
\label{8}
\epsilon ({\rm div } {\bf E}_a-gf_{abc}{\bf A}_b {\bf E}_c)=\rho _a, \;\;\;\;
{\rm curl } {\bf B}_a-\epsilon \frac {\partial {\bf E}_a}{\partial t} -
gf_{abc}(\epsilon \Phi _b{\bf E}_c + [{\bf A}_b{\bf B}_c])={\bf j}_a.
\end{equation}
The space-time dispersion of $\epsilon $ is neglected here.
 
In terms of potentials these equations are cast in the form
\begin{eqnarray}
\bigtriangleup {\bf A}_a-\epsilon \frac{\partial ^2{\bf A}_a}{\partial t^2}=
-{\bf j}_a -
gf_{abc}(\frac {1}{2} {\rm curl } [{\bf A}_b, {\bf A}_c]+\frac {\partial }
{\partial t}({\bf A}_b\Phi _c)+[{\bf A}_b {\rm curl } {\bf A}_c]-  \nonumber \\
\epsilon \Phi _b\frac 
{\partial {\bf A}_c}{\partial t}- 
\epsilon \Phi _b {\rm grad } \Phi _c-\frac {1}{2} gf_{cmn}
[{\bf A}_b[{\bf A}_m{\bf A}_n]]+g\epsilon f_{cmn}\Phi _b{\bf A}_m\Phi _n), \hfill \label{f.6}
\end{eqnarray}

\begin{eqnarray}
\bigtriangleup \Phi _a-\epsilon \frac {\partial ^2 \Phi _a}
{\partial t^2}=-\frac {\rho _a}{\epsilon }+ 
gf_{abc}(2{\bf A}_c {\rm grad }\Phi _b+{\bf A}_b
\frac {\partial {\bf A}_c}{\partial t}+ 
\frac {\partial \Phi _b}{\partial t}
\Phi _c)-  \nonumber  \\
g^2 f_{amn} f_{nlb} {\bf A}_m {\bf A}_l \Phi _b. \hfill  \label{f.7}
\end{eqnarray}
If the terms with coupling constant $g$ are omitted, one gets
the set of Abelian equations, that differ from electrodynamical equations
by the color index $a$ only. The external current is due to a parton  
moving fast relative to partons "at rest".

The crucial distinction between (\ref{f.2}) and  (\ref{f.6}),
(\ref{f.7}) is that there is no radiation (the field strength is zero in
the forward light-cone and no gluons are produced) in the lowest order solution 
of (\ref{f.2}), and it is admitted for (\ref{f.6}), (\ref{f.7}),
because $\epsilon $ takes into account the collective response (color
polarization) of the nuclear matter. 

Cherenkov effects are especially suited for treating them by classical
approach to (\ref{f.6}), (\ref{f.7}). Their unique feature is
independence of the coherence of subsequent emissions on the time interval
between these processes. The lack of balance of the phase $\Delta \phi $ between
emissions with frequency $\omega =k/\sqrt {\epsilon }$ separated by the
time interval $\Delta t $ (or the length $\Delta z=v\Delta t$) is given by
\begin{equation}
\label{f.9}
\Delta \phi =\omega \Delta t-k\Delta z\cos \theta =
k\Delta z(\frac {1}{v\sqrt {\epsilon }}-\cos \theta )
\end{equation}
up to terms that vanish for large distances. For Cherenkov effects the 
angle $\theta $ is
\begin{equation}
\label{f.10}
\cos \theta = \frac {1}{v\sqrt {\epsilon }}.
\end{equation}
The coherence condition $\Delta \phi =0$ is valid independent of $\Delta z $.
This is a crucial property specific for Cherenkov radiation only.
The fields $(\Phi _a, {\bf A}_a)$ and the classical current for in-medium 
gluodynamics can be represented by the product of the 
electrodynamical expressions $(\Phi , {\bf A})$ and the color matrix $T_a$. 

Let us recall the Abelian solution for the current 
with velocity ${\bf v}$ along $z$-axis:
\begin{equation}
\label{f.11}
{\bf j}({\bf r},t)={\bf v}\rho ({\bf r},t)=4\pi g{\bf v}\delta({\bf r}-{\bf v}t).
\end{equation}

In the lowest order the solutions for the scalar and vector potentials are
related 
${\bf A}^{(1)}({\bf r},t)=\epsilon {\bf v} \Phi ^{(1)}({\bf r},t)$ and
\begin{equation}
\label{f.12}
\Phi ^{(1)}({\bf r},t)=\frac {2g}{\epsilon }\frac {\theta
(vt-z-r_{\perp }\sqrt {\epsilon v^2-1})}{\sqrt {(vt-z)^2-r_{\perp} ^2
(\epsilon v^2-1)}}.
\end{equation}

Here $r_{\perp }=\sqrt {x^2+y^2}$ is the cylindrical coordinate; $z$
symmetry axis. The cone
\begin{equation}
\label{f.14}
z=vt-r_{\perp }\sqrt {\epsilon v^2-1}
\end{equation}
determines the position of the shock wave due to the $\theta $-function
in (\ref{f.12}). The field is localized within this cone and decreases with time 
as $1/t$ at any fixed point. The gluons emission is perpendicular to the cone
(\ref{f.14}) at the Cherenkov angle (\ref{f.10}). 

Due to the antisymmetry of $f_{abc}$, the higher order terms ($g^3$,...)
are equal to zero for any solution multiplicative in space-time and color
as seen from (\ref{f.6}), (\ref{f.7}).

The expression for the intensity of the radiation is given by the Tamm-Frank
formula (up to Casimir operators) that leads to infinity for constant 
$\epsilon $. The $\omega $-dependence of $\epsilon $ (dispersion), its 
imaginary part (absorption) and chromomagnetic permeability can be taken
into account \cite{depj}. 

The attempts to calculate the nuclear permittivity from first principles are
not very convincing. It can be obtained from the polarization operator.
The corresponding dispersion branches have been computed in the lowest order
perturbation theory \cite{kk, we}. The properties of collective 
excitations have been studied in the framework of the thermal field
theories (see, e.g., \cite{bi}). The results with an additional 
phenomenological ad hoc assumption about the role of resonances were used
in a simplified model of scalar fields \cite{ko} to show that the nuclear 
permittivity can be larger than 1, i.e. admits Cherenkov gluons. Extensive 
studies were performed in \cite{dpri}. No final decision about the nuclear
permittivity is yet obtained from these approaches. It must be notrivial
problem because we know that, e.g., the energy dependence of the refractive 
index of water \cite{ja} (especially, its imaginary part) is so complicated 
that it is not described quantitatively in electrodynamics.  

Therefore, we prefer to use the general formulae of the scattering theory to 
estimate the nuclear permittivity. It is related to the refractive index $n$ 
of the medium $\epsilon =n^2$
and the latter one is expressed through the real part of the forward 
scattering amplitude of the refracted quanta ${\rm Re}F(0^{\rm o},E)$ by
\begin{equation}
\label{f.19}
{\rm Re} n(E )=1+\Delta n_R =1+\frac {6m_{\pi }^3\nu }{E^2}{\rm Re }F(E) =
1+\frac {3m_{\pi }^3\nu }{4\pi E } \sigma (E )\rho (E ).  
\end{equation}
Here $E$ denotes the energy, $\nu $ the number of scatterers within a single 
nucleon, $m_{\pi }$ the pion mass, $\sigma (E)$ the cross section and $\rho (E)$ 
the ratio of real to imaginary parts of the forward scattering amplitude $F(E)$. 

Thus the emission of Cherenkov gluons is possible only for processes with 
positive ${\rm Re} F(E)$ or $\rho (E)$. Unfortunately, we are unable to 
calculate directly in QCD these characteristics of gluons
and have to rely on analogies and our knowledge of the properties of hadrons. 
The only experimental facts we get for this medium are brought about 
by particles registered at the final 
stage. They have some features in common, which (one may hope!) are also
relevant for gluons as the carriers of the strong forces. Those are the resonant
behavior of amplitudes at rather low energies and the positive real part of the
forward scattering amplitudes at very high energies for hadron-hadron and
photon-hadron processes as measured from the interference of the Coulomb and
hadronic parts of the amplitudes. ${\rm Re} F(0^{\rm o},E)$ is always positive 
(i.e., $n>1$) within the low-mass wings of the Breit-Wigner resonances.  
This shows that the necessary condition for Cherenkov effects $n>1$ is
satisfied at least within these two energy intervals. This fact was used
to describe experimental observations at SPS, RHIC and cosmic ray energies. 
The asymmetry of
the $\rho $-meson shape at SPS \cite{da} and azimuthal correlations of 
in-medium jets at RHIC \cite{ul, ajit} were explained by emission of 
comparatively low-energy Cherenkov gluons \cite{dnec, drem1}.
The parton density and intensity of the radiation were estimated. In its turn,
cosmic ray data \cite{apan} at energies corresponding to LHC require very 
high-energy gluons to be emitted by the ultrarelativistic partons moving along 
the collision axis \cite{d1}. Let us note the important difference from
electrodynamics, where $n<1$ at high frequencies. 

The in-medium equations are not Lorentz-invariant. There is no problem in
macroscopic electrodynamics, because the rest system of the macroscopic matter
is well defined and its permittivity is considered there. For collisions of
two nuclei (or hadrons) it requires special discussion.

Let us consider a particular parton that radiates in the nuclear matter.
It would "feel" the surrounding medium at rest if the momenta of all other 
partons, with which this parton can interact, are smaller and sum to zero. 
In RHIC experiments the triggers, that registered the jets (created by partons), 
were positioned at 90$^{\rm o}$ to the collision axis. Such partons should be
produced by two initial forward-backward moving partons scattered at 90$^{\rm o}$.
The total momentum of the other partons (medium spectators) is balanced, because 
for such a geometry the partons from both nuclei play the role of spectators 
forming the medium. Thus the center
of mass system is the proper one to consider the nuclear matter at rest in
this experiment. The permittivity must be defined there. The Cherenkov rings 
consisting of hadrons have been registered around the away-side jet, which
traversed the nuclear medium. This geometry requires, however, high statistics,
because the rare process of scattering at 90$^{\rm o}$ has been chosen. 

The forward 
(backward) moving partons are much more numerous and have higher energies. 
However, one cannot treat the radiation of such a primary parton in the c.m.s.
in a similar way, because the momentum of the spectators is different from zero,
i.e. the matter is not at rest. Now the spectators (the medium) are formed
from the partons of another nucleus only. Then the rest system of the medium
coincides with the rest system of that nucleus and the permittivity should 
refer to this system. The Cherenkov radiation of such highly energetic 
partons must be considered there. That is what was done for interpretation
of the cosmic ray event in \cite{d1}. This discussion shows that 
one must carefully define the rest system for other geometries of the 
experiment with triggers positioned at different angles.

Thus our conclusion is that the definition of $\epsilon $ depends on the 
geometry of the experiment. Its corollary is that partons moving in different
directions with different energies can "feel" different states of matter in
the ${\bf same}$ collision of two nuclei because of the permittivity dispersion. 
The transversely scattered partons with comparatively 
low energies can analyze the matter with rather large permittivity corresponding
to the resonance region, while the forward moving partons with high energies 
would "observe" a low permittivity in the same collision. This peculiar feature
can help scan the $(\ln x, Q^2)$-plane as discussed in \cite {drem2}.
It explains also the different values of $\epsilon $ needed for the description
of the RHIC and cosmic ray data.

 These conclusions can be checked at LHC, because 
both RHIC and cosmic ray geometry will become available there. The energy
of the forward moving partons would exceed the thresholds above which 
$n>1$. Then both types of experiments can be done, i.e. the 90$^{\rm o}$-trigger
and non-trigger forward-backward partons experiments. The predicted results for 
90$^{\rm o}$-trigger geometry are similar to those at RHIC. The non-trigger
Cherenkov gluons should be emitted within the rings at polar angles of tens
degrees in c.m.s. at LHC by the forward moving partons (and symmetrically by 
the backward ones) according to some events observed in cosmic
rays \cite{apan, drem1}.

Let us compare the conclusions for Cherenkov and Mach
shock waves. The Cherenkov gluons are described as the transverse waves
while the Mach waves are longitudinal. Up to now, no experimental signatures
of these features were proposed. 

The most important experimental fact is the
position of the maxima of humps in two-particle correlations. They are
displaced from the away-side jet by 1.05-1.23 radian \cite{st, fw, ph}.
This requires rather large values of ${\rm Re} \epsilon \sim 2-3$ and indicates
high density of the medium \cite{drem1} that agrees with other conclusions.
The fits of the humps with complex permittivity are in progress.
The maxima due to Mach shock waves should be shifted by the smaller value
0.955 if the relativistic equation of state is used ($\cos \theta = 1/\sqrt 3$).
To fit experimental values one must consider different equation of state.
In three-particle correlations, this displacement is about 1.38 \cite{ul}. 

There are some claims \cite{ul, ajit} that Cherenkov effect contradicts to 
experimental observations because it predicts the shift of these maxima
to smaller angles for larger momenta. They refer to the prediction made in
\cite{ko}. However, the conclusions of this paper about the momentum 
dependence of the refractive index can hardly be considered as
quantitative ones because the oversimplified  scalar $\Phi ^3$-model with
simplest resonance insertions was used 
for computing the refractive index. In view of difficult task of its
calculation discussed above, the fits of maxima seem to be more important
for our conclusions about the validity of the two schemes.     

Mach waves should appear for forward moving partons at RHIC but were not
found. The energy threshold of $\epsilon $ explains this phenomenon
for Cherenkov gluons.

Work supported by RFBR grants 06-02-17051, 06-02-16864, 08-02-91000-CERN.

\end{document}